\documentclass[aps,prl,twocolumn,nofootinbib,amsmath,amssymb,floatfix,10pt]{revtex4-1}

\usepackage{graphicx,color,amsmath,amsxtra}
\usepackage{amssymb}
\usepackage[unicode]{hyperref}
\usepackage{graphicx}
\usepackage{epstopdf}

\begin{document}

\title{Gravitational waves from oscillons in a generalized exponential plateau potential}

\author{Peter Lott}
\email{peter.lott@phenikaa-uni.edu.vn}
\affiliation{Phenikaa Institute for Advanced Study, Phenikaa University, Hanoi 12116, Vietnam}

\author{Tuan Q. Do}
\email{tuan.doquoc@phenikaa-uni.edu.vn}
\affiliation{Phenikaa Institute for Advanced Study, Phenikaa University, Hanoi 12116, Vietnam}

\date{\today}

\begin{abstract}
We study oscillon formation and gravitational wave production in a generalized
exponential plateau inflationary potential. Using Floquet analysis, we identify
parametric instability bands consistent with oscillon formation, and solve
numerically for the oscillon profile, finding quasi-breather solutions with
lifetimes $\tau_{\rm osc} \cdot m_{\rm eff} \sim 10^3$ --- $6\times10^4$.
Applying the poltergeist mechanism, we compute the induced gravitational wave
spectrum and find a peak at $f_{\rm peak} \approx 2.5\times10^{10}$~Hz with
amplitude $\Omega_{\rm GW,0}\,h^2 \sim 10^{-9}$--$10^{-8}$ for the benchmark
parameter $\beta_{\rm pot} = 5\times10^{-6}$. This is far below the region
forbidden by big bang nucleosynthesis. For $\beta_{\rm pot} = 5\times10^{-5}$
the signal exceeds the big bang nucleosynthesis bound for all considered
oscillon energy fractions, constraining the parameter space of the model. The
signal falls in the GHz regime, potentially accessible to future resonant
cavity experiments.
\end{abstract}

\maketitle

\section{Introduction}

The post-inflationary universe is a rich arena for nonlinear scalar field
dynamics. After the end of slow-roll inflation, the inflaton condensate
oscillates around the minimum of its potential and can fragment into long-lived,
spatially localized configurations known as \textit{oscillons}
\cite{Bogolyubsky:1976nx, Gleiser:1993pt, Copeland:1995fq}. These objects form
through parametric resonance driven by the anharmonic structure of the inflaton
potential \cite{Kofman:1994rk, Kofman:1997yn, Greene:1997fu}, and can persist
for many orders of magnitude longer than a single oscillation period
\cite{Gleiser:2008ty, Amin:2010jq, Amin:2011hj, Salmi:2012ta}. Their formation
is generic for a broad class of plateau-like potentials satisfying
$M^2 \ll M_{\rm Pl}^2$ \cite{Amin:2010xe, Amin:2010dc, Lozanov:2017hjm}.

The observational consequences of oscillons are severalfold. If they dominate
the post-inflationary energy budget, their eventual decay sources a burst of
induced gravitational waves via the poltergeist mechanism
\cite{Inomata:2019ivs, p16t-xz3k} --- a resonant amplification of scalar
perturbations triggered by the sudden transition from oscillon-dominated matter
domination to radiation domination. This mechanism is directly analogous to the
enhanced GW production in primordial black hole reheating scenarios
\cite{Inomata:2020lmk}, and generically produces signals in the MHz--GHz
frequency range, potentially accessible to future resonant cavity experiments
based on the inverse Gertsenshtein effect \cite{Aggarwal:2020olq, Berlin:2021txa}.

In a recent systematic study \cite{p16t-xz3k}, the poltergeist GW
spectrum was computed for several well-known inflationary potentials, including
$\alpha$-attractor T-models, axion monodromy, and hilltop models. A key finding
is that the GW spectrum encodes direct information about the inflaton potential
through the oscillon mass, lifetime, and formation scale, offering a novel
observational window into the inflationary epoch complementary to current and
upcoming CMB measurements, including recent constraints from ACT DR6
\cite{ACT:2025fju, ACT:2025tim}.

In this Letter we extend this programme to a new inflationary potential with a
generalized exponential plateau, recently proposed in \cite{Kouniatalis:2025orn}.
This potential is distinguished by an exceptionally flat plateau leading to
extra-suppressed tensor-to-scalar ratios $r \lesssim 10^{-5}$, well within
current observational bounds \cite{Planck:2018vyg}. We show that for
appropriate choices of the transition scale $M$, the potential satisfies the
conditions for oscillon formation, and we solve numerically for the oscillon
profile, finding quasi-breather solutions with lifetimes
$\tau_{\rm osc} \cdot m_{\rm eff} \sim 10^3$--$6\times10^4$. Applying the
poltergeist formalism of \cite{p16t-xz3k}, we compute the induced GW
spectrum and find a peak at $f_{\rm peak} \approx 2.5\times10^{10}$~Hz. For
the benchmark parameter $\beta_{\rm pot} = 5\times10^{-6}$, the signal lies
below the BBN bound \cite{Maggiore:1999vm, Cyburt:2015mya} for all oscillon
energy fractions considered, while $\beta_{\rm pot} = 5\times10^{-5}$ is ruled
out, providing a direct constraint on the potential parameter space from GW
physics.

\section{Potential and Parameter Choices}

\begin{figure}[h]
    \includegraphics[width=0.95\linewidth]{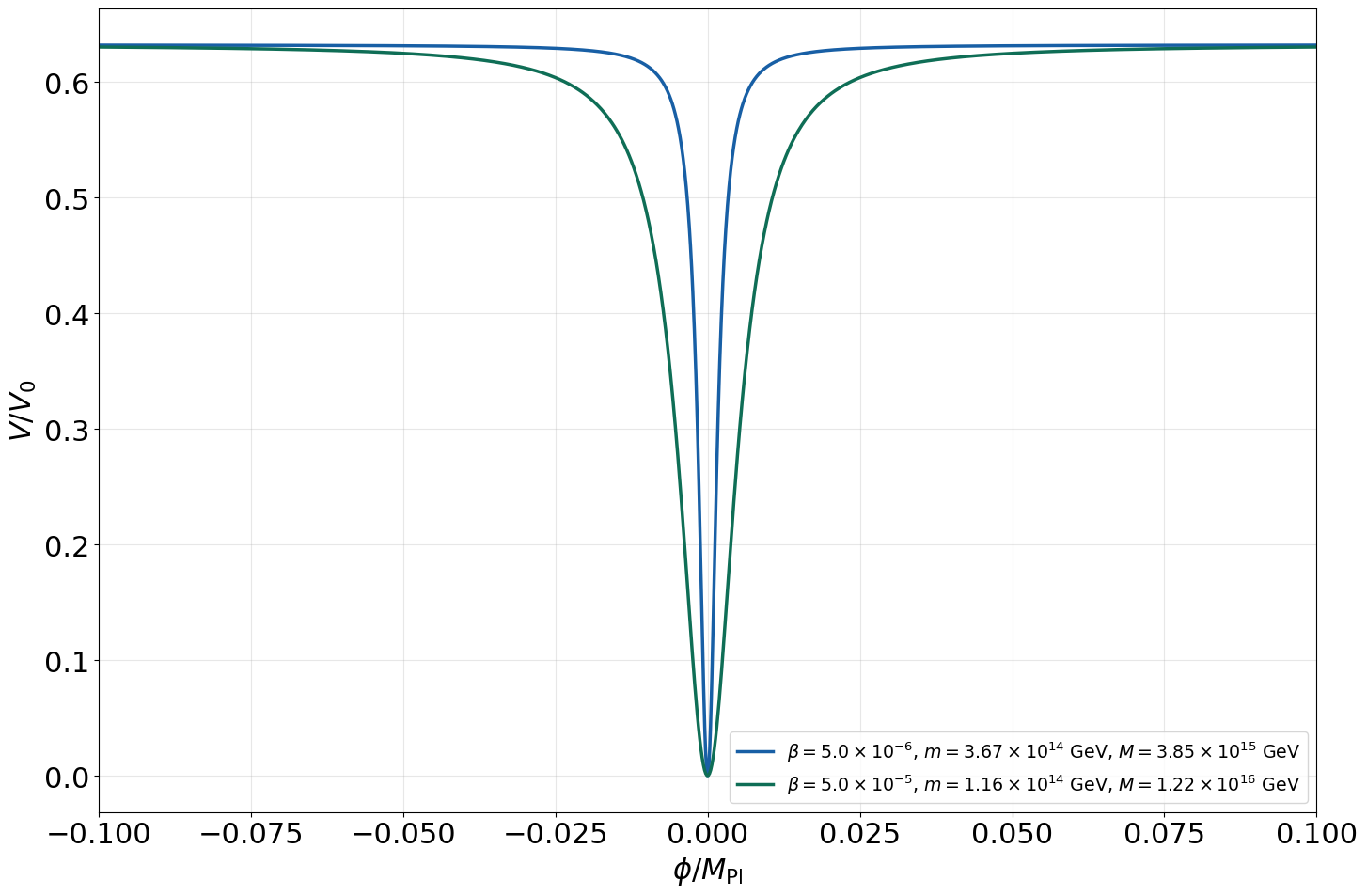}
    \caption{The generalized exponential plateau potential
    $V(\phi) = m_{\rm eff}^2 M^2(1 - e^{-F(\phi)})$ in units of
    $V_0 = m_{\rm eff}^2 M^2$, plotted as a function of $\phi/M$ for
    $\beta_{\rm pot} = 5\times10^{-5}$, $\alpha = \gamma = 1$. The potential
    exhibits a quadratic minimum at the origin and saturates to a plateau
    $V \to V_0$ at large field values. The transition between the two regimes
    occurs at $\phi \sim M$, which sets the scale for oscillon formation.}
    \label{fig:potential}
\end{figure}

We consider the generalized exponential plateau potential \cite{Kouniatalis:2025orn}
\begin{equation}
V(\phi) = m_{\rm eff}^2 M^2 \left(1 - e^{-F(\phi)}\right),
\label{eq:potential}
\end{equation}
where
\begin{equation}
F(\phi) = \frac{\bar\alpha(\phi/M)^2}{\bar\beta + \bar\gamma(\phi/M)^2},
\end{equation}
with $m_{\rm eff}^2 M^2 = V_0$, $\bar\alpha = \alpha M^2$,
$\bar\beta = \beta_{\rm pot} M_{\rm Pl}^2$, and $\bar\gamma = \gamma M^2$.
Here $\alpha$ $\beta_{\rm pot}$ and $\gamma$ are dimensionless parameters. Note that $\beta_{\rm pot}$ is
distinct from the oscillon energy fraction $\beta_{\rm osc}$ introduced in
later in this Letter. Near the minimum, Eq.~\eqref{eq:potential} reduces to
\begin{equation}
V(\phi \ll M) \approx \tfrac{1}{2}m_{\rm eff}^2\phi^2
- \tfrac{3}{8}\frac{m_{\rm eff}^2}{M^2}\phi^4 + \mathcal{O}(\phi^6),
\label{eq:quartic}
\end{equation}
which holds for $\alpha = \gamma = 1$ and $\beta_{\rm pot} \ll 1$, satisfying
the conditions for oscillon formation \cite{p16t-xz3k} with $c_g = 0$
and $c_\lambda = 3/8$.

The effective inflaton mass $m_{\rm eff}$ is defined by the curvature of the
potential at the origin,
\begin{equation}
m_{\rm eff}^2 \equiv V''(0) = \frac{2\alpha}{\beta_{\rm pot}}\,\frac{V_0}{M_{\rm Pl}^2},
\label{eq:meff}
\end{equation}
which follows from Eq.~\eqref{eq:potential} using
$\bar\alpha = \alpha M^2$ and $\bar\beta = \beta_{\rm pot} M_{\rm Pl}^2$.
Since $V_0$ and $\alpha$ are fixed by inflationary observables,
Eq.~\eqref{eq:meff} determines $m_{\rm eff}$ directly from $\beta_{\rm pot}$,
independently of $M$. The transition scale $M$ is then chosen
self-consistently by imposing $\lambda \equiv V_0/(m_{\rm eff}^2 M^2) = 1$,
ensuring nonlinear effects enter at order unity, which gives
\begin{equation}
M = \frac{\sqrt{V_0}}{m_{\rm eff}}.
\label{eq:Mscale}
\end{equation}
A single choice of $\beta_{\rm pot}$ therefore uniquely fixes both
$m_{\rm eff}$ and $M$ via Eqs.~\eqref{eq:meff}--\eqref{eq:Mscale}. Smaller
$\beta_{\rm pot}$ yields a larger $m_{\rm eff}$ and hence a smaller $M$,
i.e.\ a larger ratio $m_{\rm eff}/M$, as verified from Table~\ref{tab:params}.
The hierarchy $m_{\rm eff}^2 \ll M^2 \ll M_{\rm Pl}^2$ required for classical
oscillon formation \cite{p16t-xz3k} is satisfied for both benchmark
values considered.

The inflationary observables fix $V_0 = 2\times10^{60}$~GeV$^4$ and the
inflationary value $\beta \sim 1.7\times10^{-7}$ \cite{Kouniatalis:2025orn}.
We study oscillon formation for
$\beta_{\rm pot} \in \{5\times10^{-6}, 5\times10^{-5}\}$, both larger than
the inflationary $\beta$, with self-consistent mass scales summarized in
Table~\ref{tab:params}.

\begin{table}[h]
\centering
\begin{tabular}{ccc}
\hline\hline
$\beta_{\rm pot}$ & $m_{\rm eff}$ (GeV) & $M$ (GeV) \\
\hline
$5\times10^{-6}$ & $3.7\times10^{14}$ & $3.9\times10^{15}$ \\
$5\times10^{-5}$ & $1.2\times10^{14}$ & $1.2\times10^{16}$ \\
\hline\hline
\end{tabular}
\caption{Benchmark parameters satisfying 
$m_{\rm eff}^2 \ll M^2 \ll M_{\rm Pl}^2$, obtained from $\beta_{\rm pot}$
via Eqs.~\eqref{eq:meff}--\eqref{eq:Mscale}.}
\label{tab:params}
\end{table}

\section{Floquet Analysis}

To assess whether the potential supports oscillon formation, we study
parametric instabilities of the post-inflation oscillating condensate~\cite{Amin:2011hj,Lozanov:2017hjm}.
Perturbing the homogeneous background as
$\phi(t,\vec{x}) = \phi_0(t) + \delta\phi(t,\vec{x})$ and linearizing the
Klein-Gordon equation (neglecting Hubble friction, justified for
$H \ll m_{\rm eff}$), each Fourier mode $\delta\phi_k(t)$ satisfies the
Hill equation
\begin{equation}
\delta\ddot\phi_k + \left[k^2 + V''(\phi_0(t))\right]\delta\phi_k = 0.
\label{eq:hill}
\end{equation}
Here $V''(\phi_0(t)) \equiv d^2V/d\phi^2|_{\phi=\phi_0(t)}$ is the second
derivative of the potential evaluated along the homogeneous background
trajectory $\phi_0(t)$, which satisfies $\ddot\phi_0 + V'(\phi_0) = 0$
and is periodic with period $T = 2\pi/m_{\rm eff}$.
By Floquet's theorem, solutions of Eq.~\eqref{eq:hill} take the form
$\delta\phi_k(t) = e^{\mu_k t} P_k(t)$ with $P_k$ periodic, and modes with
${\rm Re}(\mu_k) > 0$ grow exponentially. We compute $\mu_k$ numerically from
the transfer matrix $\mathcal{M}$, obtained by evolving two linearly
independent solutions of Eq.~\eqref{eq:hill} over one period $T$, via~\footnote{Since Eq.~\eqref{eq:hill} has no first-derivative term, Abel's
theorem gives $\det\mathcal M=1$, so the eigenvalues of $\mathcal M$
satisfy $\lambda^2-(\mathrm{tr}\,\mathcal M)\lambda+1=0$, i.e.
$\lambda_{\max}=\tfrac{1}{2}\big(|\mathrm{tr}\,\mathcal M|+\sqrt{(\mathrm{tr}\,\mathcal M)^2-4}\big)$.
With $\lambda=e^{\mu_k T}$ and the identity
$\mathrm{arccosh}(x)=\ln(x+\sqrt{x^2-1})$, this yields
Eq.~\eqref{eq:floquet} directly~\cite{Teschl:2012ode,Yakubovich:1975periodic}.}
\begin{equation}
\mu_k = \frac{1}{T}\,{\rm arccosh}\!\left(\frac{|{\rm tr}\,\mathcal{M}|}{2}\right),
\label{eq:floquet}
\end{equation}
valid whenever $|{\rm tr}\,\mathcal{M}| > 2$; otherwise $\mu_k = 0$.
Explicitly, $\mathcal{M}$ is the $2\times2$ matrix whose columns are the
values $(\delta\phi_k(T), \delta\dot{\phi}_k(T))^\top$ for the two initial
conditions $(\delta\phi_k(0), \delta\dot{\phi}_k(0))^\top = (1,0)$ and
$(0,1)$ respectively. The
resulting instability bands confirm that modes with $k \sim m_{\rm eff}$ are
exponentially amplified, consistent with oscillon formation.

\section{Oscillon Profile and Lifetime}

We seek time-periodic, spatially localized solutions of the form
\begin{equation}
\phi(r,t)=\Phi(r)\cos(\omega t),
\end{equation}
with $\omega<m_{\rm eff}$, where $m_{\rm eff}$ is defined in
Eq.~\eqref{eq:meff}. Here $\phi(r,t)$ is the scalar field, $\Phi(r)$ is its
radial profile, and $\omega$ is the oscillation frequency. Introducing the
dimensionless variables
\begin{equation}
\rho=m_{\rm eff}r,\qquad
\varphi(\rho)=\frac{\Phi(r)}{M},
\end{equation}
the radial profile satisfies
\begin{equation}
\varphi''+\frac{2}{\rho}\varphi'
+\left(\frac{\omega}{m_{\rm eff}}\right)^2\varphi
-\frac{1}{m_{\rm eff}^2M}
\left.\frac{dV}{d\phi}\right|_{\phi=M\varphi}=0,
\label{eq:oscillon_ode}
\end{equation}
where primes on $\varphi$ denote derivatives with respect to $\rho$. The
boundary conditions are regularity at the origin, $\varphi'(0)=0$, and the
exponential tail condition
\begin{equation}
\varphi'(\rho)+\kappa\varphi(\rho)\to0,
\qquad
\kappa=\sqrt{1-\left(\frac{\omega}{m_{\rm eff}}\right)^2},
\end{equation}
as $\rho\to\infty$.

We solve Eq.~\eqref{eq:oscillon_ode} numerically using a shooting method,
scanning $\omega/m_{\rm eff}\in[0.50,0.97]$ and determining the central
amplitude $\varphi_0$ with Brent's method to satisfy the asymptotic boundary
condition. The resulting solutions exhibit a small-amplitude oscillatory
radiation tail, identifying them as \textit{quasi-breathers}, consistent with
the general expectation that exact breathers do not exist in $3+1$ dimensions
\cite{Segur:1987mg,Fodor:2008es}.

The radiation tail amplitude $\varphi_{\rm tail}/\varphi_0$ determines the
scaling of the energy loss rate and hence the oscillon lifetime,
\begin{equation}
\tau_{\rm osc}\sim
\frac{1}{(\varphi_{\rm tail}/\varphi_0)^2\,m_{\rm eff}}.
\label{eq:tauosc}
\end{equation}
Scanning the quasi-breather family, we find lifetimes
$\tau_{\rm osc}m_{\rm eff}\sim6\times10^3$ near
$\omega/m_{\rm eff}=0.97$, rising to
$\tau_{\rm osc}m_{\rm eff}\approx5.7\times10^4$ near
$\omega/m_{\rm eff}=0.73$. We adopt
$\tau_{\rm osc}m_{\rm eff}=5.7\times10^4$ as our benchmark value.

Assuming a fraction $\beta_{\rm osc}\in(1/2,1)$ of the inflaton energy density
is transferred into oscillons at formation, the initial oscillon number density
is
\begin{equation}
n_{\rm osc}=\frac{\beta_{\rm osc}\,\rho_{\rm inf}}{m_{\rm osc}},
\end{equation}
where $\rho_{\rm inf}=3H_f^2M_{\rm Pl}^2$ and
$m_{\rm osc}=\alpha_m M^2/m_{\rm eff}$ with $\alpha_m\simeq100$
\cite{p16t-xz3k}.

\section{Gravitational Wave Spectrum}

If $\beta_{\rm osc} \sim \mathcal{O}(1)$, oscillons briefly dominate the
energy density, driving a transient matter-dominated phase whose decay
sources induced gravitational waves via the poltergeist mechanism
\cite{p16t-xz3k,Inomata:2019ivs}. The poltergeist formula assumes
sudden decay; our quasi-breathers instead radiate continuously, which
spreads the same energy release over a longer interval and suppresses the
coherent buildup of the induced spectrum, so Eq.~(\ref{eq:omegapeak}) gives
an upper bound on the true signal.

Following \cite{p16t-xz3k}, the resonant spectrum peaks at
\begin{equation}
\begin{gathered}
\Omega_{\rm GW,res}(k) \approx \Omega_{\rm GW}^{\rm peak}
\left(\frac{k}{k_{\rm osc}}\right)^5\Theta_{\rm uv}(k),
\\[4pt]
\Omega_{\rm GW}^{\rm peak} = \frac{C^2}{3456\pi}\,c_s(1-c_s^2)^2
\left(\frac{k_f}{k_{\rm rh}}\right)^{7}\frac{k_f}{k_{\rm osc}}
\cdot\frac{m_{\rm osc}}{4\pi M_{\rm Pl}}\left(\frac{H_f}{M_{\rm Pl}}\right)^{1/3}
\end{gathered}
\label{eq:omegapeak}
\end{equation}
with IR tail
\begin{equation}
\Omega_{\rm GW,IR}(k) \approx \frac{c_s^4 C^2}{324\pi^2}
\left(\frac{k_f}{k_{\rm osc}}\right)^{8}
\left(\frac{k_{\rm osc}}{k_{\rm rh}}\right)^{6}
\frac{k}{k_{\rm osc}},
\label{eq:omegaIR}
\end{equation}
where $c_s=1/\sqrt3$, $k_f=H_f$, $k_{\rm rh}=\tau_{\rm osc}^{-1}$,
$k_{\rm osc}\approx m_{\rm eff}$, $\Theta_{\rm uv}$ cuts off the spectrum
above $k_{\rm osc}$, and $C \equiv C(\beta_{\rm osc})$ is given in \cite{p16t-xz3k}. Redshifting to today, $\Omega_{\rm GW,0}h^2 \approx 0.387\,(g_*/106.75)^{-1/3}\Omega_{r,0}h^2\,
\Omega_{\rm GW,RD}$, with peak frequency
$f_{\rm peak} = (m_{\rm eff}/2\pi)(T_0/T_{\rm rh})(g_{*s,0}/g_{*s,\rm rh})^{1/3}$,
assuming Standard Model degrees of freedom $g_*=g_{*s}=106.75$ at $T_{\rm rh}$~\cite{Husdal:2016haa}.

\begin{figure}[h]
\centering
\includegraphics[width=\linewidth]{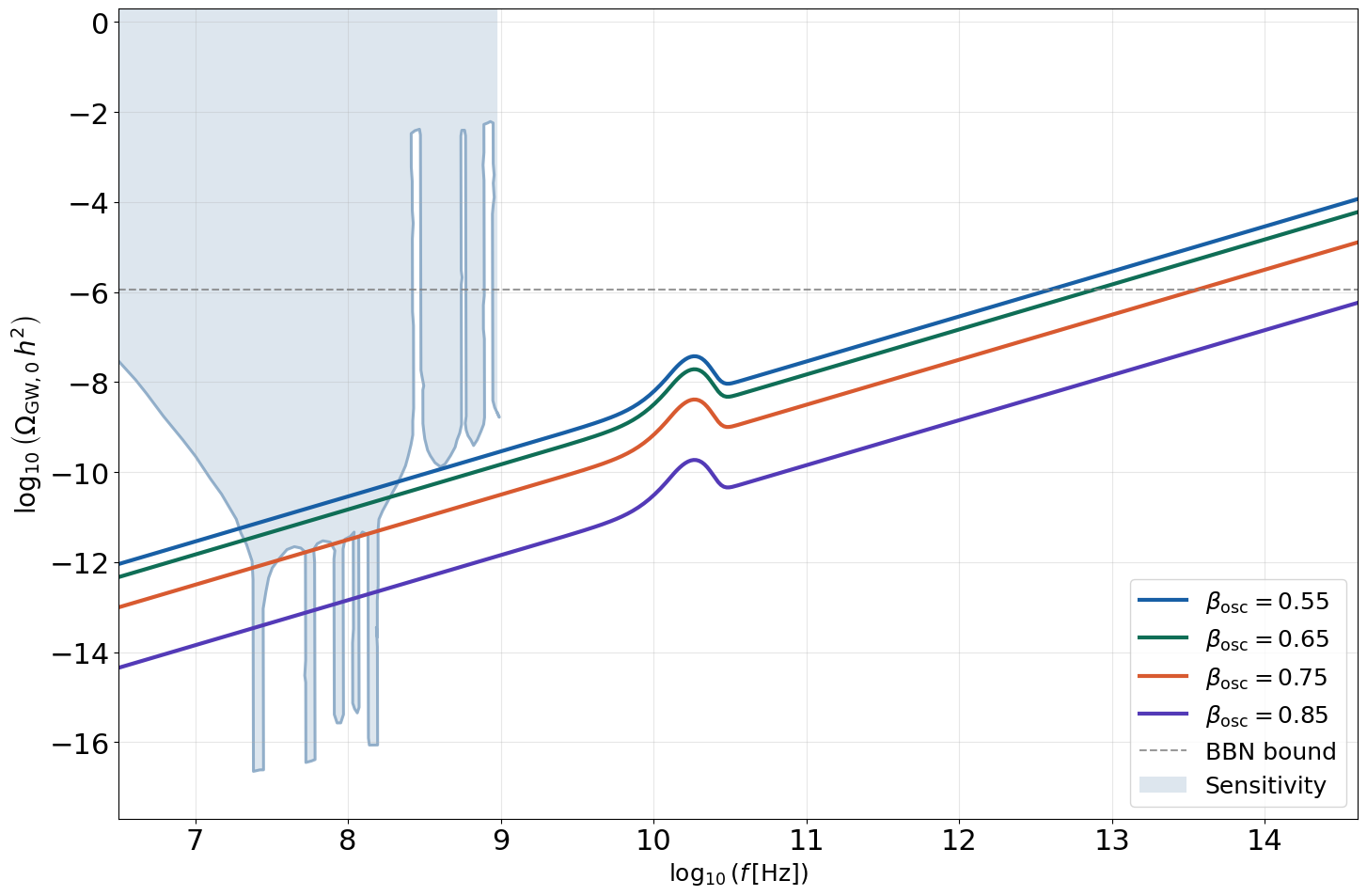}
\caption{Poltergeist GW spectrum for $\beta_{\rm pot} = 5\times10^{-6}$ and
$\tau_{\rm osc}\cdot m_{\rm eff} = 5.7\times10^4$, for $\beta_{\rm osc} \in
\{0.55, 0.65, 0.75, 0.85\}$ (top to bottom). The dashed line is the BBN
bound. All curves are below the bound. The peak at
$f_{\rm peak} \approx 2.5\times10^{10}$~Hz falls in the GHz regime targeted
by future resonant cavity experiments \cite{Aggarwal:2020olq, Berlin:2021txa}. Sensitivity curve digitized from Fig. 2 of \cite{p16t-xz3k}.}
\label{fig:gw}
\end{figure}

For $\beta_{\rm osc}\in[0.55,0.85]$ (Fig.~\ref{fig:gw}), $\beta_{\rm pot}
=5\times10^{-6}$ gives $f_{\rm peak}\approx2.5\times10^{10}$~Hz and
$\Omega_{\rm GW,0}h^2 \sim 10^{-9}$--$10^{-8}$, safely below the BBN bound
$\Omega_{\rm GW,0}h^2 \lesssim 1.12\times10^{-6}$ \cite{Maggiore:1999vm} for
all $\beta_{\rm osc}$ considered. $\beta_{\rm pot}=5\times10^{-5}$ exceeds
this bound for all $\beta_{\rm osc}$, ruling out that combination.
For $\beta_{\rm pot}\geq5\times10^{-4}$, $\Omega_{\rm GW,IR}$ itself exceeds
unity, signaling breakdown of the perturbative poltergeist treatment.

\section{Conclusions}

We have investigated oscillon formation and gravitational wave production in
the generalized exponential plateau potential proposed in
\cite{Kouniatalis:2025orn}. This potential is characterized by an
exceptionally flat plateau leading to extra-suppressed tensor-to-scalar ratios,
and admits a compatible form for oscillon formation when the transition scale
$M$ satisfies $m_{\rm eff}^2 \ll M^2 \ll M_{\rm Pl}^2$.

Using Floquet analysis we identified parametric instability bands consistent
with oscillon formation for benchmark parameters
$\beta_{\rm pot} \in \{5\times10^{-6}, 5\times10^{-5}\}$ with
self-consistent mass scales satisfying
$\lambda \equiv V_0/(m_{\rm eff}^2 M^2) = 1$. We then solved numerically for
the oscillon profile using a shooting method, finding a family of
quasi-breather solutions --- spatially localized configurations with a small
but nonzero radiating tail, consistent with the general expectation that exact
breathers do not exist in $3+1$ dimensions \cite{Segur:1987mg, Fodor:2008es}.
By scanning the quasi-breather family as a function of $\omega/m_{\rm eff}$,
we identified a longest-lived solution with
$\tau_{\rm osc} \cdot m_{\rm eff} \approx 5.7\times10^4$, which we adopted as
a benchmark for the gravitational wave calculation.

Applying the poltergeist formalism of \cite{p16t-xz3k}, we computed the
induced gravitational wave spectrum for the two benchmark values of
$\beta_{\rm pot}$. For $\beta_{\rm pot} = 5\times10^{-6}$ we find a peak at
$f_{\rm peak} \approx 2.5\times10^{10}$~Hz with amplitude
$\Omega_{\rm GW,0}\,h^2 \sim 10^{-9}$--$10^{-8}$, safely below the BBN bound
\cite{Maggiore:1999vm, Cyburt:2015mya} for all oscillon energy fractions
$\beta_{\rm osc} \in [0.55, 0.85]$. For $\beta_{\rm pot} = 5\times10^{-5}$,
the peak amplitude exceeds the BBN bound for all values of $\beta_{\rm osc}$
considered, ruling out this parameter combination and providing a direct
constraint on the potential parameter space from gravitational wave physics.
For $\beta_{\rm pot} \geq 5\times10^{-4}$, the IR tail of the poltergeist
formula exceeds unity, indicating that the approximation breaks down and a more
careful treatment is required.

We note several caveats. Our quasi-breathers radiate continuously rather than
decaying suddenly, so the poltergeist formula --which assumes a sharp
matter-to-radiation transition -- provides an upper bound on the signal rather
than an exact prediction. The oscillon energy fraction $\beta_{\rm osc}$ and
lifetime $\tau_{\rm osc}$ are treated as free parameters here; their precise
values require lattice simulations with CosmoLattice \cite{Figueroa:2021yhd},
which we leave for future work. Such simulations would sharpen the GW
prediction to a single curve rather than a band, and would validate or refine
the quasi-breather picture found here analytically.

The allowed signal for $\beta_{\rm pot} = 5\times10^{-6}$ peaks in the GHz
frequency range, potentially accessible to future resonant cavity experiments
\cite{Aggarwal:2020olq, Berlin:2021txa, Herman:2022fau} operating via the
inverse Gertsenshtein effect. This places the model's GW signature in a
frequency window complementary to existing and planned interferometric
detectors, and highlights the potential of high-frequency GW searches as a
probe of inflationary model building.

\begin{acknowledgments}
PL acknowledges NAFOSTED funding under grant number 103.01-2025.147. TQD is funded by the Vietnam National Foundation for Science and Technology Development (NAFOSTED) under grant
number 103.01-2023.50. 
\end{acknowledgments}

\bibliographystyle{apsrev4-1}
\bibliography{reference}

\end{document}